\documentstyle[11pt,fake_vol,twoside,epsf]{article}
\begin{document}

\title{The Large Area Lyman Alpha Survey}
 \author{James E. Rhoads}
\affil{Space Telescope Science Institute, 3700 San Martin Drive, Baltimore, 
MD 21218 USA}
\author{Sangeeta Malhotra}
\affil{Johns Hopkins University}
\author{Arjun Dey \& Buell T. Jannuzi}
\affil{National Optical Astronomy Observatory}
\author{Daniel Stern}
\affil{NASA Jet Propulsion Laboratory}
\author{Hyron Spinrad}
\affil{University of California, Berkeley}

\begin{abstract}
The Lyman-$\alpha$ line is expected to be strong in the presence of active star
formation and the absence of dust, making it a good tool for finding
chemically primitive galaxies in the early universe.  We report on a
new survey for high redshift Lyman-$\alpha$ sources, the Large Area Lyman Alpha
(LALA) survey.  Our survey achieves an unprecedented combination of
volume and sensitivity by using narrow-band filters on the new
$8192^2$ pixel CCD Mosaic Camera at the 4 meter Mayall telescope of
Kitt Peak National Observatory.
 
Well-detected sources with flux and equivalent width matching known
high redshift Lyman-$\alpha$ galaxies have an observed surface density
corresponding to $11000 \pm 700$ per square degree per unit redshift
at $z=4.5$.  Early spectroscopic followup from the Keck telescope
suggests that $\sim 1/3$ of these are actually at $z\approx 4.5$, and
has confirmed five $z > 4$ Lyman-$\alpha$ emitters so far.  Combining our
photometric survey with spectroscopic results, we estimate a net
density of $\sim 4000 $ Lyman-$\alpha$ emitters per square degree per unit
redshift at $z\approx 4.5$.  The star formation rate density
(estimated both from UV continuum and from line emission) is
comparable to that of the Lyman break galaxy population within present
uncertainties.  The most extreme Lyman-$\alpha$ emitters in our sample have
rest frame equivalent widths $> 100\AA$, consistent with the
expectations for the first burst of star formation in a primitive,
dust-free galaxy.
\end{abstract}

\section{Introduction}
The first major burst of star formation in a young, high redshift
galaxy is expected to produce copious numbers of ionizing photons
thanks to an abundance of hot, massive main sequence stars.  The
interstellar medium that gave birth to the stars can then convert 2/3
of these photons into Lyman-$\alpha$ photons as neutral hydrogen atoms are
ionized and recombine.  The strong Lyman-$\alpha$ line emission that can result
was first proposed as a signpost of young galaxies in formation by
Partridge \& Peebles (1967; hereafter PP67).

However, searches for Lyman-$\alpha$ emission at high redshift failed to
discover a field population of such galaxies for thirty years.
(See review by Pritchet 1994 and references therein; and more
recently, Thompson \& Djorgovski 1995; Thommes et al.~1998.)
These searches placed stringent upper limits on the abundance and
brightness of the Lyman-$\alpha$ line at high redshift.  More recent searches
have finally found Lyman-$\alpha$ emitters in the field
(Cowie \& Hu 1998; Hu, Cowie \& McMahon 1998; Pascarelle, Windhorst,
\& Keel 1998; Dey et al.~1998; Hu, McMahon, \& Cowie 1999; Steidel et
al.~2000; Kudritzki et al.~2000; Manning et al.~2000),
but at line luminosities $\sim 10^{-2}$ below the early predictions by
PP67.

Several factors could be responsible for this paucity of bright Lyman-$\alpha$
sources, including dust effects, small star formation units, and
long star formation time scales.  (PP67 based their estimates on dust
free, Milky Way-sized galaxies that form most of their stars in
$3\times 10^7$ years.)

Dust can quench this line very effectively: Lyman-$\alpha$ photons are
resonantly scattered by neutral hydrogen atoms, and so traverse a much
longer path than continuum photons do in order to escape the galaxy
where they are emitted.  This long path enhances the optical depth for
line photons, reducing the line equivalent width.  Quantitative
estimates of this effect are however quite sensitive to the spatial
distribution of the ISM (e.g. Neufeld 1991).  In contrast, star
formation in small units would yield small line and continuum
luminosities while leaving equivalent widths unaffected.  Under this
scenario, the expected numbers of Lyman-$\alpha$ sources would rise in inverse
proportion to their luminosity.  Protracted star formation would also
reduce both line and continuum luminosities.  Additionally, a star
formation event lasting longer than the lifetime of a massive star
would allow metals and dust produced by the first generation of stars
to pollute the galaxy's ISM, resulting again in dust extinction and
potential reductions in Lyman-$\alpha$ equivalent width.

Good statistical samples of Lyman-$\alpha$ emitters that yield distributions of
luminosities and equivalent widths should allow us to distinguish
among the above possibilities.  We are conducting the Large Area Lyman
Alpha (LALA) survey to obtain such a sample.

Whatever the reason for the faintness of Lyman-$\alpha$ sources, the effects of
dust provide another strong reason for seeking these objects: Truly
primordial galaxies, taken here to mean chemically unevolved systems,
will have large Lyman-$\alpha$ equivalent widths because of their low dust
content.  Lyman-$\alpha$ surveys should therefore preferentially select for
this most interesting class of high-redshift galaxy.  This is not to
say that all Lyman-$\alpha$-selected galaxies will be primordial, nor even that
a large fraction will be.  Dustier galaxies can have substantial
Lyman-$\alpha$ equivalent widths if their ISM is patchy, or if it is in bulk
motion (Kunth et al 1998).  However, if chemically primitive galaxies
are to be found at all, the Lyman-$\alpha$ search is a good tool to discover
them.

\section{The Survey}
The Large Area Lyman Alpha (LALA) survey is possible because of new
wide field instrumentation.  We exploit the 8192$^2$ pixel CCD Mosaic
camera at the Kitt Peak National Observatory's 4m Mayall telescope to
achieve a survey efficiency (measured in the product $A \Omega$ of
collecting area and solid angle) some 6 times greater than comparably
deep narrowband surveys at Keck.  The ``core'' survey uses five
overlapping $80$\AA\ bandpass filters ($\lambda_c = 6559$, $6611$,
$6650$, $6692$, and $6730 $\AA) to look for Lyman-$\alpha$ line emission at
$4.37 < z < 4.57$.  The corresponding survey volume is about $8.5
\times 10^5$ comoving Mpc$^3$ per 36$'$ field for $H_0 = 70 {\rm\,km\,s^{-1}\,Mpc^{-1}}$,
$\Omega=0.2$, $\Lambda=0$.  We integrate for about 6 hours per filter
per field, and achieve a limiting sensitivity $\sim 2 \times 10^{-17}
 {\rm\,erg\,cm^{-2}\,s^{-1}}$ in the line.  Further details, including a brief description
of data reduction, are given in Rhoads et al (2000).

Additionally, we use supporting broad band images from the NOAO Deep
Widefield Survey (B$_w$, R, and I filters; see Schommer, Jannuzi, \&
Dey 2001, this meeting) and from our own data (V and SDSS z' filters),
and have recently obtained $8200$\AA\ narrowband data to extend the
LALA survey to $z \approx 5.7$.

Spectroscopic followup of some LALA candidates has been carried out
using the Keck Observatory's Low-Resolution Imaging Spectrograph.  To
date, we have confirmed five emission line galaxies at $z>4$.  Two
representative spectra are shown in figures~1 and~2.  One of these two
is a galaxy with very little continuum that would have gone undetected
in a broadband survey of even extreme depth, while the other is a
less spectacular emission line source with a clearly detected
continuum and Lyman break.

\begin{figure}
\plotfiddle{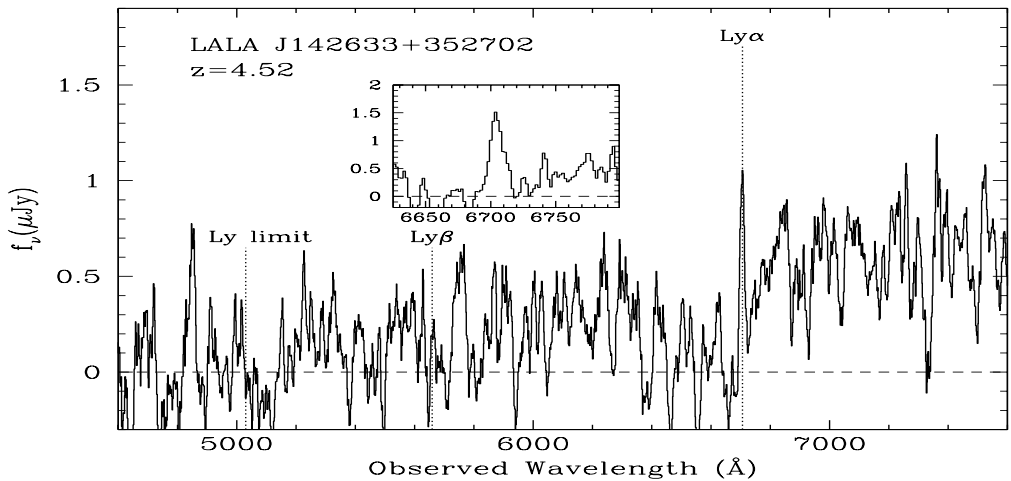}{2.in}{-90}{52}{33}{-220}{185}
\caption{
Keck spectrum of a confirmed $z=4.52$ Lyman-$\alpha$ source.  This object has a
line flux of $1.7 \times 10^{-17}  {\rm\,erg\,cm^{-2}\,s^{-1}}$ and a rest frame
equivalent width of $15$\AA\.  The line is asymmetric (see inset) and
has a strong continuum decrement from the red to the blue side, both
of which are expected for high-redshift Lyman-$\alpha$ emitters (Stern \&
Spinrad 1999).  In order to accentuate the well-sampled line and
suppress noise, the spectrum of this source has been smoothed with a
boxcar filter of width 9 pixels $= 16.6$\AA\ [main panel] or 3 pixels
$= 5.5$\AA\ [inset panel].}
\end{figure}

\begin{figure}
\plotone{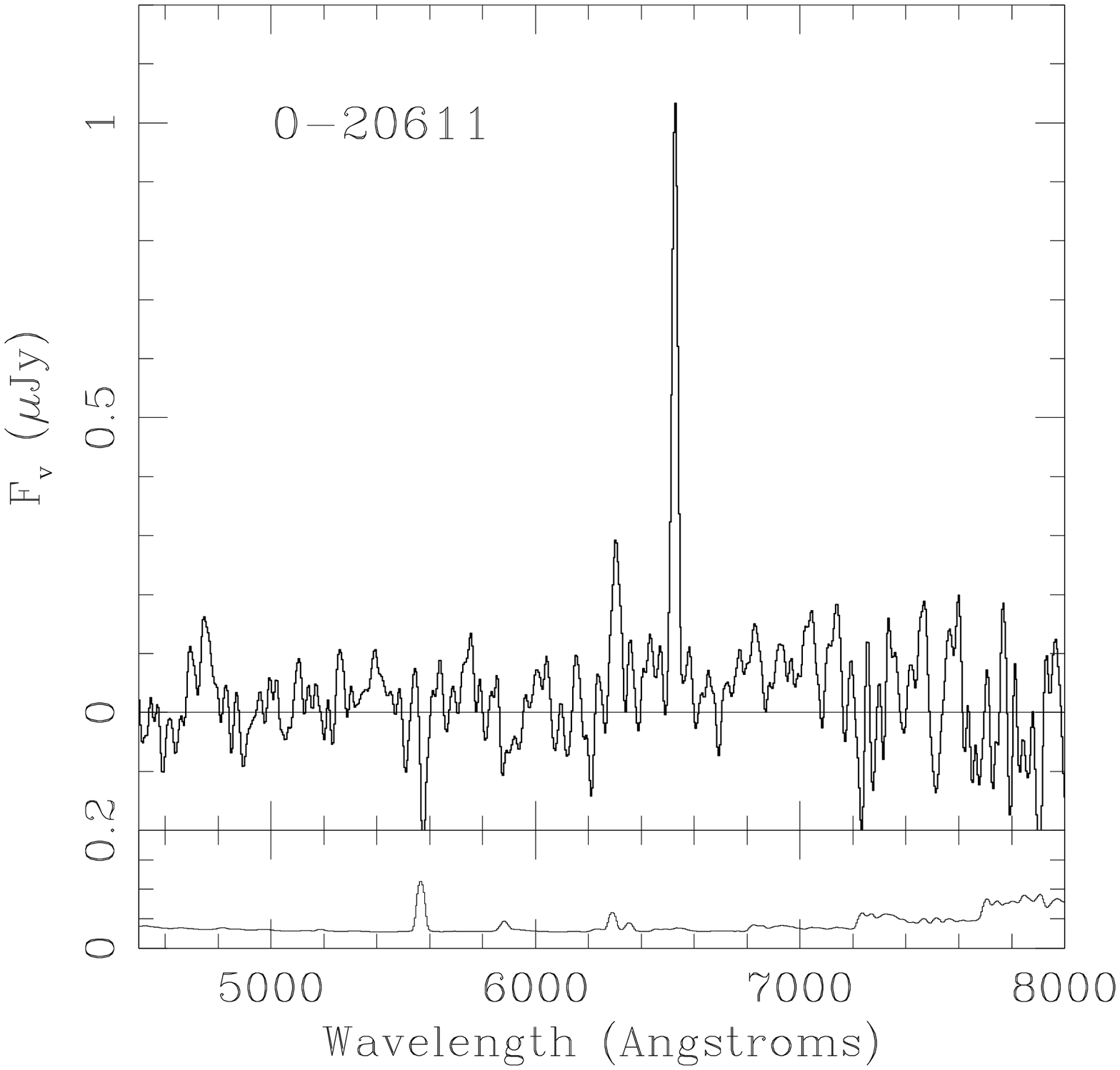}
\caption{ Keck spectrum of another confirmed Lyman-$\alpha$ source, this time
at $z=4.37$.  This object has a line flux of $\sim 4 \times 10^{-17}
{\rm\,erg\,cm^{-2}\,s^{-1}}$ and a rest frame equivalent width $> 100 \AA$ at the $2
\sigma$ level.  The continuum emission from this source is undetected
in our deep broadband images and at best marginally detected in the
spectrum.  In order to accentuate the well-sampled line and
suppress noise, the spectrum of this source has been smoothed with a
Gaussian filter with a 4 pixel $=19.4$\AA\ full width at half
maximum.  The lower panel shows $1\sigma$ photon counting errors,
suitably adjusted for the smoothing.  The weak ``line'' to the left of
Lyman-$\alpha$ is a residual from the subtraction of the $6300$\AA\ sky line.}
\end{figure}

\section{The Lyman-$\alpha$ Source Population}
Based on our photometric results and on the statistics of our first
spectroscopic followup, we inferred a source density of $\sim 4000$ Lyman-$\alpha$
emitters per square degree per unit redshift at $z=4.5$, with an uncertainty
dominated by small numbers of spectra (Rhoads et al 2000).  

We can estimate the star formation rate density in the Lyman-$\alpha$ galaxies
in two quasi-independent ways.  First, we coadd the observed $z'$ band
flux of all our good candidates and apply starburst models to convert
this rest-frame UV continuum into a star formation rate.  Correcting
this rate for our estimate that $1/3$ of the good candidates are
actually $z=4.5$ Lyman-$\alpha$ emitters, we obtain a rate of $13 \times
10^{-3} M_\odot \hbox{yr}^{-1} \hbox{Mpc}^{-3}$.  Second, we coadd the
continuum-corrected line fluxes of all our good candidates and apply a
standard conversion factor to go from Lyman-$\alpha$ flux to ionizing
luminosity and thence star formation rate.  Applying again a $1/3$
correction factor to account for foreground contamination, we find a
lower value of $2.2 \times 10^{-3} M_\odot \hbox{yr}^{-1} \hbox{Mpc}^{-3}$.  The
difference in these two estimates may be due to the effects of dust,
which would affect the Lyman-$\alpha$ based estimate more strongly than the UV
continuum estimate.  Alternatively, our simple correction factors of
$1/3$ for foreground contamination could be wrong if the $z\approx
4.5$ sources have a systematically stronger line to continuum ratio
than the remaining candidates.  While these star formation rates are
still quite uncertain (pending further spectroscopy), they are
comparable to the Lyman break galaxy rate at $z=4$, which is $5 \times
10^{-3} M_\odot \hbox{yr}^{-1} \hbox{Mpc}^{-3}$ (Madau et al 1996).

The spectra in figures~1 and~2 are representative of
our spectroscopically confirmed $z>4$ galaxies in the sense that about
half have rest frame equivalent widths $> 100$\AA, which is near the
upper limit for star forming galaxies ($100$ to $200$\AA;
Charlot \& Fall 1993).  This suggests that some of the LALA sources
may indeed be dust-free and chemically primitive.  Further
spectroscopy to rule out weak narrow-lined AGNs and/or galactic winds
(see Kunth et al 1998) will help confirm this possibility. 

\section{Future Directions}
Photometric redshift constraints derived from our suite of broadband
filters, combined with consistency checks from our growing
spectroscopic sample, will enable us to use the entire LALA sample to
study population properties of the Lyman-$\alpha$ emitters.  These include the
luminosity function, distribution of equivalent widths, and clustering
properties.  Planned near-infrared followup of selected sources will
allow us to confirm the Lyman-$\alpha$ line identification through an {[O II]
$\lambda$3727} line search and to look for older stellar populations
through a study of rest-frame optical light.  Finally, searches in new
narrowband filters will allow us to study the evolution of the Lyman-$\alpha$
source population.  Ultimately, we hope to identify statistical
samples of primitive galaxies at a range of redshifts from $z=4.5$ to
$z>6$.

\acknowledgements
We thank Andy Bunker and Steve Dawson for help with the spectroscopic
observations, and the IRAF team for
writing and helping with the MSCRED package.  JER's research is
supported by an STScI Institute Fellowship.  SM's research is
supported by NASA through Hubble Fellowship grant \# HF-01111.01-98A
from the Space Telescope Science Institute, which is operated by the
Association of Universities for Research in Astronomy, Inc., under
NASA contract NAS5-26555.  The data presented herein were obtained at
the Mayall Telescope of the Kitt Peak National Observatory; and at the
W.M. Keck Observatory, which is operated as a scientific partnership
among the California Institute of Technology, the University of
California and the National Aeronautics and Space Administration.  The
Keck Observatory was made possible by the generous financial support
of the W.M. Keck Foundation.


\end{document}